# DIGITAL TRANSFORMATION OF URBAN PLANNING IN AUSTRALIA: INFLUENCING FACTORS AND KEY CHALLENGES

Soheil Sabri[a,1] and Sherah Kurnia[b]

[a] Urban Digital Twin Lab, School of Modeling Simulation and Training, University of Central Florida, Email:

soheil.sabri@ucf.edu

[b] Department of Computing and Information Systems. The University of Melbourne, Email: sherahk@unimelb.edu.au

**Abstract**

Over the past two decades, several governments in developing and developed countries have started their journey toward digital transformation. However, the pace and maturity of digital technologies and strategies are different between public services. Current literature indicates that research on the digital transformation of urban planning is still developing. Therefore, the aim of this study is to understand the influencing factors and key challenges for the digital transformation of urban planning in Australia. The study adopts the inter-organisational theory and Planning Support Science (PSScience) under the Technological, Organisational, and External Environmental (TOE) framework. It involves a multiple case study administrered semi-structured interviews with thirteen IT and urban planning experts across Victoria and New South Wales governments and private industries. The study findings indicate that the main challenges for digital transformation of the Australian urban planning system are related to organisational and external environmental factors. Further- more, a digital maturity model is absent in the Australian urban planning industry. This study offers important implications to research and practice related to digital transformation in urban planning.

*Keywords:* Digital Transformation, ePlanning, PSScience, Urban Planning, TOE framework

---

[1] Corresponding Author

## 1. Introduction

Recently, digital transformation in the public service sector has received unprecedented attention from industry and academia (Batty, 2021; Gong et al., 2020; ITU, 2019). However, this transformation's pace and maturity level are not equal among all services. For example, the fast-growing areas of digitalisation are health care provision, tax service provision, and election and citizen participation services (ITU, 2019). Therefore, digital transformation in urban planning remained in the early stages (Batty, 2021).

Urban planning (or simply "planning") is a multi-disciplinary activity that creates better places and public spaces for people by balancing the built and natural environment, addressing community needs, respecting cultural significance, and considering sustainable prosperity of the urban and regions (Planning Institute of Australia, 2021b). Some of the disciplines that planners involve are housing, energy, transport, and tourism, development, and city strategies. In Australia, planning tasks include preparing plans and detailing their application methods, which are strategic in nature. In addition, planners implement plans via the regulation of development, called statutory planning or development assessment (Burton, 2017). Therefore, in this study, the term planning refers to plan-making, statutory planning, strategic planning, and development assessment tasks (Williams et al., 2012).

Current literature acknowledges the significant effect of digital transformation in modifying and improving policy-design procedures in cities and regions (Panori et al., 2021). In addition, innovations offer new opportunities through emerging technologies such as digital twins (Dembski et al., 2020; Tzachor et al., 2022), big data analytics (Talebkhah et al., 2021), artificial intelligence (Boje et al., 2020), and IoT sensor data (Sabri, 2021b) for better planning and policy-making. Public services deal with complex, multi-layer, and multi-actor regulatory environments. The planning complexities might hinder the progression of information systems application in spatial planning tasks (Antwi Acheampong et al., 2021). For example, in 2011, the Australian National eDevelopment Assessment (eDA) steering committee suggested that electronic tools in planning process can integrate different stakeholders, harmonise fragmented online services and increase planning business efficiencies. As such, an ePlanning roadmap was proposed to fully digitise the planning process in all jurisdictions in the country by 2020 (National eDA Steering Committee, 2011). However, the roadmap failed to meet its two primary goals:

- By 2015 all jurisdictions will implement at least essential ePlanning services (Basic use of technology in the planning and development services)
- By 2020 all jurisdictions will implement premium ePlanning services (Extensive use of technology for planning and real-time information

streamlining for maximum utility users)

The literature suggests that digital transformation is not a technical process, but institutional and governance, business requirements, and internal and external factors play vital roles (Gong et al., 2020). While computational modelling and geospatial technologies have been applied in urban planning tasks for more than six decades, there is little study on digital infrastructures that facilitate the interaction of planners with the associated community and other stakeholders. In addition, research on spatial data infrastructures (SDIs), which have been implemented in many planning applications, indicates the vital role of geospatial data along with legal, policy, and institutional domains (Crompvoets et al., 2008; Schindler et al., 2018). (Rajabifard et al., 2016) demonstrated how SDI harmonises urban data and facilitates multi-disciplinary collaborations in urban planning. (Chen et al., 2020) demonstrated the value of SDIs in deriving urban indicators to measure and monitor city performance. However, there is little research on leveraging opportunities such as SDI frameworks in planning digital reform and understanding the success or failure factors. Furthermore, the adoption of these technologies remained conceptual, and little empirical study is available on the practical and functional digital transformation of planning activities. This research aims at understanding the influencing factors and key challenges for the digital transformation of urban planning in Australia.

To achieve this aim, this study intends to develop a conceptual framework defining influencing factors and key challenges in digital transformation strategies of planning and map them to the complexities of planning processes in Australia. The research question addressed is:

*What are the influencing factors and key challenges in the digital transformation of Australian urban planning?*

Given that the Australian planning rules are different in each state, this exploratory study employs semi-structured interviews with 13 experts in urban planning from the two most populous states of Australia. The study findings indicate that there is a lack of coordination between different organisations' information systems and the return value of digital transformation in urban planning is not well conceived. Therefore, the limitation in alignment between IT and planning (as a business) is one of the major issues in the digital transformation of planning in Australia.

This paper is structured as follows. In the next section, we provide a literature review with a focus on inter-organisational collaborations, diffusion of innovation, institutional theory, and the application of information systems in the planning process, which is called Planning Support Science (PSSicence). We will also propose a conceptual framework to conduct this exploratory research in two Australian States, Victoria and NSW. Then we will explain the methods of data collection and analysis in the next section, which will be proceeded by the report on results. Next, we provide an explanation of the major findings and discuss the contradictions and supporting findings with the previous studies in section 5. Finally, in section 6, we provide the theoretical and practical implications of our findings in the digital transformation of planning in Australia. We also suggest future work to improve the current study.

## 2. Literature Review

Urban planning is a multi-layer and multi-actor activity, and information and tasks are shared in organisations horizontally (between departments) and vertically (between local, state, and federal levels). Therefore, in this study, an inquiry of theories related to the concept of Inter-organisational Information Systems (IOS) is essential (de Corbière et al., 2019). Accordingly, three other associated scientific areas were identified: Planning Support Science (PSScience) (Geertman & Stillwell, 2020), Diffusion of Innovation (DOI) (Rogers, 1995), and Institutional Theory (IT) (Teo et al., 2003). Given that this study specifically emphasises on the application of information systems in urban planning, we adopted the PSScience, which is an emerging research stream, but there are enough scientific works that make it possible to be explored from theoretical, methodological, and empirical aspects.

A relatively small body of literature is concerned with the digital transformation of urban planning. While a large and growing body of literature has investigated the digital transformation in public services (Gong et al., 2020; ITU, 2019) and its role in economic growth (Mićić, 2017; Tan et al., 2022), the dialogue for investigation of this phenomenon in urban planning just started (Batty, 2021). It is important to make a distinction between the studies emphasised on the planning support tools and this study. We are aware that there is growing literature about urban analytics infrastructure (Rajabifard et al., 2019), decision support, and planning tools (Amirebrahimi et al., 2016). The literature either focuses on a specific planning process such as sustainable urban change (de Jesus Dionisio et al., 2020; Dembski et al., 2020) or spatial planning and design outcomes (Langenheim et al., 2022; Schindler et al., 2020; Schindler & Dionisio, 2021). However, the focus of this study is on the bigger picture and the role of planning as a business and its relationship with the organisational and inter-organisational information systems and business strategies.

*Inter-organisational Collaboration*

Geertman (2006) argued that there is a mismatch between the application of planning technologies and their outcomes in the planning for the supply and demand of urban services. Therefore, he suggested a framework, which adds other dimensions to the technology. Geertman's (2006) framework considers the content and structure of planning tasks, users' characteristics, the specifics of the planning and policy process, the political context, and the dominant planning style and policy model. (Goodspeed & Hackel, 2017) used Geertman's (2006) framework to understand the benefits and challenges of rolling out an innovative planning support system (PSS) in Southern California. Furthermore, they argued that with the anticipation of adopting the PSS by other sections in an organisation, theories of diffusion of innovation (Rogers, 1995) and organisational restructuring (Ventura, 1995) need to be included in their study. Goodspeed and Hackel's (2017) findings suggest the importance of inter-organisational collaboration and access to the infrastructure, users' participation in the design and implementation process, and embracing the value of new technologies. Their findings align with the institutional theory, suggesting that organisations with close relationships and data exchange adopt a similar structure, even though its inefficiency is unknown (de Corbière et al., 2019).

However, recent studies highlighted the challenge of lacking a collaboration culture within and between organisations (Biderman & Swiatek, 2020; Evans et al., 2019). Planners are in contact with many other stakeholders, including architects, environmental engineers, transport experts, and structure engineers. In addition, the government planning process is a multi-organisational and collaborative task (Afzali et al., 2014). Therefore, it is essential to understand the external factors influencing the organisational decisions for change (Teo et al., 2003).

A number of authors have considered two external factors that influence the organisational strategies for change: political (Biderman & Swiatek, 2020; Tomor & Geertman, 2020) and regulatory support (Musakwa & Moyo, 2020). Evans et al. (2019) found that the digital transformation pace is accelerated by solid political agenda in the Australian public services. However, political support is not enough unless there is an action toward regulatory changes (Daniel & Pettit, 2021).

One of the critical influencing factors for the digital transformation of governments is the citizen-centric aspect (ITU, 2019). In the context of urban planning, the users' participation in the design and implementation of the PSS reflects the citizen-centric viewpoint. The findings of Musakwa and Moyo (2020) in Africa and (S. Wang et al., 2020) in China for the application of the PSS in ePlanning have also reflected the essential role of adopting citizen-centric technologies. Moreover, Christensen & Lægreid, (2020) highlighted the lack of government-to-citizens tools, which has implications for perceived coordination quality and success of the digital transformation.

From embracing the value of new technologies perspective, suggested by Goodspeed and Hackel (2017), recent literature highlighted two potential challenges in the adoption of digital technologies in planning. First, researchers found that the value of digital innovations for planning is not recognised across organisations (Daniel, 2020). Second, due to the lack of capacity, scarcity of resources, and perceived cost of technology, organisations don't consider urban planning as a priority in their digital transformation and IT development strategies (Biderman & Swiatek, 2020; Criado-Perez et al., 2021; Geertman & Stillwell, 2020; Panori et al., 2021; Tomor & Geertman, 2020).

*Planning Support Science (PSScience)*

Recently, Geertman and Stillwell (2020), alluded to a paradigm shift toward the emergence of Planning Support Science (PSScience) based on recent innovations and digital applications in planning tasks worldwide. They elaborated the PSScience in three dimensions: application, governance, and instrumentation.

From an application perspective, it is suggested that planning innovations have been moving from sustainability, resilience, and city futures to new frontiers such as business intelligence and behavioural analytics. However, the new boundaries might be challenged by the availability of valuable data. The literature highlighted that data availability, accessibility, and reliability remained current issues (Musakwa & Moyo, 2020; Sabri et al., 2016; Tomor & Geertman, 2020). In addition, the lack of planning data standards, which was discussed by Agius et al. (2018), Daniel (2020) [32], and Criado-Perez et al. (2021) challenges the previous and new planning application frontiers. Furthermore, the complexity of the planning process is reported to be a barrier in many jurisdictions (Agius et al., 2018; S. Wang et al., 2020).

The second dimension refers to spatial planning governance, which emphasises multi-organisational collaboration in planning tasks. Geertman and Stillwell (2020) believe that spatial planning is a collaborative environment where the public and private sectors, civil society, and non-profit organisations work together. However, one of the biggest challenges is to test different forms of knowledge and synchronise additional requirements in this environment. Moreover, as indicated before, the culture of inter-organisational collaboration is another challenge reported in the recent literature (Daniel Pettit, 2021). Besides, the capacity to accept and adapt to digital technologies differ among organisations, which hinders the collaborations among different organisations (Criado-Perez et al., 2021; Panori et al., 2021).

The third dimension is instrumentation, which refers to the technical aspect of PSScience. The planning community has admitted that information technology and smart cities have changed how public and private services are delivered (Damurski, 2018). However, the challenge will be whether a related skillset already exists (Agius et al., 2018; Musakwa & Moyo, 2020) and how the technological innovations can be adopted in social innovations and change society's culture and behaviour (Sabri, 2021a). Furthermore, the literature concerns digital literacy among urban planners. Several recent studies indicated the lack of capability to develop and use 3D data and open-source tools for more advanced analytics among planners (Agius et al., 2018; Geertman & Stillwell, 2020; Musakwa & Moyo, 2020). From the digital infrastructure's implementation point of view, the recent studies consistently reported the issue of silo systems (Anthony Jnr, 2021; Rajabifard et al., 2016), lack of interoperability, technology standards (Agius et al., 2018; Criado-Perez et al., 2021), and scarcity of available open-source tools (Pettit et al., 2020).

*Knowledge Gap*

Table 1 summarises the most influential factors and challenges for digital transformation in urban planning extracted from the literature. These studies indicate a relationship between digital transformation of planning and public and private organisations' digital strategies. However, it is not clear how the organisations' digital transformation journey impacted the planning process.

| Theme | Challenge | Reference |
|---|---|---|
| **Technical** | Citizen-centric platform/ user friendliness of tools | (Anthony Jnr et al., 2021; Engin et al., 2020; Moretto et al., 2022; Schito et al., 2019; Schrotter & Hürzeler, 2020; Wästberg et al., 2020) |
| | Lack of spatial and planning technology standards | (Anthony Jnr et al., 2021; Boje et al., 2020; Janečka, 2019; Meta et al., 2022; Schrotter & Hürzeler, 2020) |
| | Lack of open-source tools | (Boje et al., 2020; Moretto et al., 2022; Panori et al., 2021; Schrotter & Hürzeler, 2020) |
| | Weakness of technical support and preparation | (Moretto et al., 2022; Schrotter & Hürzeler, 2020) |
| | Systems are developed in silo | (Anthony Jnr et al., 2021; Crompvoets et al., 2008; Moretto et al., 2022; Schrotter & Hürzeler, 2020) |
| **Data** | Lack of planning data standards | (Boje et al., 2020; Breunig et al., 2020; Burton, 2017; Crompvoets et al., 2008) |
| | Data availability and usability / Open government data (OGD) | (McAuliffe & Sawyer, 2021; Schrotter & Hürzeler, 2020) |
| **Value and Cost** | Scarcity of financial capital | (Batty, 2021; Gong et al., 2020; Talebkhah et al., 2021) |
| | Overlooking the value of digital infrastructures | (Daniel, 2020; Panori et al., 2021; Tan et al., 2022) |
| | Perceived cost of technology | (Biderman & Swiatek, 2020; Panori et al., 2021; Tomor & Geertman, 2020) |
| **Planning Business** | IT-Planning strategy alignment | (Anthony Jnr et al., 2021; Planning Institute of Australia, 2021a) |
| | Planning complexity | (Bibri, 2018; Breunig et al., 2020; Planning Institute of Australia, 2021a; Sabri et al., 2019) |
| **Resources** | Digital literacy among planners | (Batty, 2021; Chassin et al., 2021; Dembski et al., 2020; Gong et al., 2020) |
| | Capability for adapting to change. Lack of relative skillset among planners | (Antwi Acheampong et al., 2021; Chassin et al., 2021) |
| **Culture** | Lack of collaboration culture | (Bibri, 2018; Biderman & Swiatek, 2020; Evans et al., 2019; Packard, 2021) |
| | Poor organisational culture Embracing the innovation | (Bibri, 2018; Packard, 2021) |

Table 1 Summary of the literature

*Conceptual Framework*

All of the studies reviewed here can be contextualised in three groups of technology, organistion, and external task environmental (TOE) framework (Tornatzky & Fleischer, 1990). The integration of technological innovation theories with the TOE framework has been widely used to understand the interaction of different factors contributing to organisation's decisions in adopting innovative technologies (Kurnia et al., 2015). Furthermore, the literature suggests that TOE can be used in any organisation as a general framework to support a systematic analysis of the factors influencing their decisions and strategies in the adoption of any technology (Kurnia et al., 2015). For example, integrating DOI and TOE in the architecture, engineering, and construction (AEC) industries supported understanding the driving forces of Building Information Modelling (BIM) adoption and its impact on inter-organisational relations (Ahuja et al., 2018; Z. Wang et al., 2021).

Therefore, three context groups of TOE framework (Technology, Organ- isation, External Task Environment) are defined as follows:

- Technology: availability and characteristics of technology in terms of its complexity for use, compatibility with existing culture and values, and feasibility of adoption in organisation with minimum risk are related to this category.

- Organisation: support by the top management, perceived costs, organisational culture, and the availability of skills are related to organisation category.

- External Task Environment: inter-organisational relations, availability of supporting infrastructure, users' demand, political willingness to adopt innovations, and regulatory support are the most critical factors related to this category.

Accordingly, a research framework is developed to inform the data collection to analyse the influencing factors and key challenges of digital transformation in Australian urban planning (Figure 1).

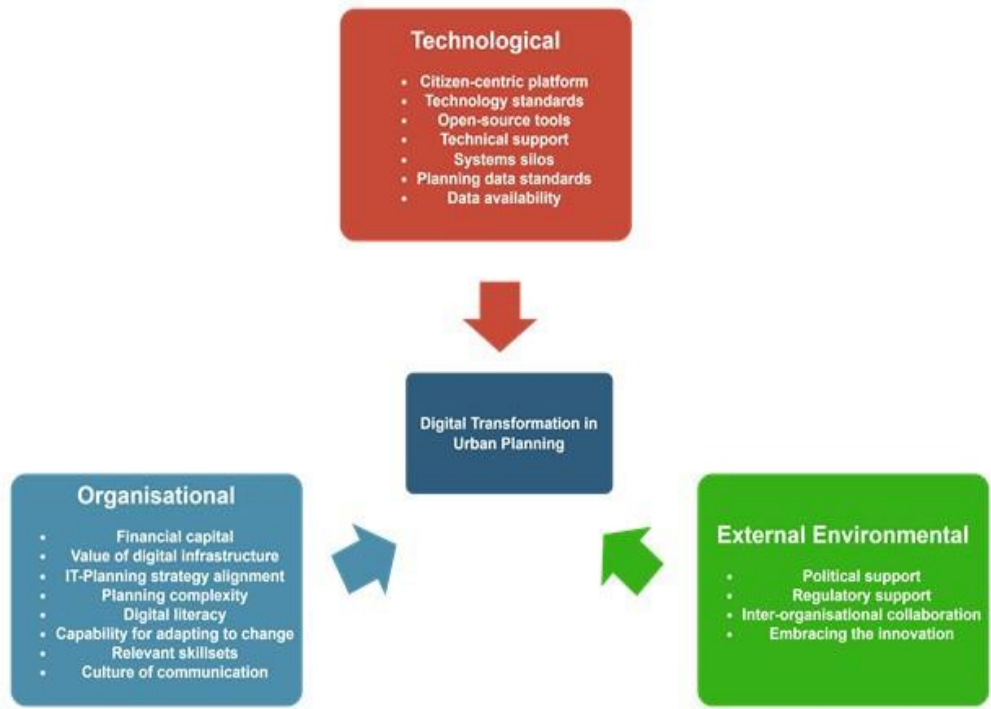

Figure 1: Conceptualising the Influencing factors and challenges for the digital transformation of urban planning

## 3. The Research Method

The study uses a multiple case study approach (Yin, 2009) to discover the influencing factors and key challenges for digital transformation within the Australian urban planning industry. The case study allows investigators to perform an in-depth examination of a phenomenon of interest (Kurnia et al., 2015), which is currently lacking in the literature on the digital transformation of urban planning. Therefore, this study contributes to the current body of knowledge, focusing on the ecosystem of multiple planning rules. According to Kurnia et al. (2015), the case study approach is appropriate for research on phenomena lacking theory. As such, given different urban planning rules in Australia, a multiple case study was selected to reflect on independent analytical conclusions, which cannot be observed in a single case study research (Yin, 2009). This argument justifies the selection of multiple case studies as the research method in this study.

To determine the ideal sample size for qualitative research involving interviews, we could not find a consensus in the literature(Francis et al., 2009) In addition, (Guest et al., 2006) recommend a minimum sample size of twelve for interview studies, as this is the number required to reach data saturation.

Thirteen respondents were selected from two states of Victoria (Vic) and New South Wales (NSW) with Urban planning, and IT background working in the government and private industries. Table 2 provides an overview of the interviewees in two states. Interviews were online, and the duration of each interview was between 45 - 60 minutes. A semi-structured and open-ended questionnaire (Appendix A) was administered, and each interview session was recorded upon receiving an officially signed consent. In some cases, follow-up conversations via phone call or email were conducted for further clarification. The interviewees were mainly from the government, but four were selected from the external organisations to capture the inter-organisational factors. The participants had different experiences, with an even distribution of mid- to above-career levels.

| Participant | Organisation | | Expertise | Work Experience (years) |
|---|---|---|---|---|
| | State | Type | | |
| P1 | Vic | Government | Solution Architect | >20 |
| P2 | Vic | Government | Automated Approvals | 10−15 |
| P3 | Vic | Government | GIS and Analytics | 10−15 |
| P4 | NSW | Peak Body Professional | Policy and Campaigns | 5−10 |
| P5 | Vic | Government | Geospatial Intelligence and Data Analyst | 5−10 |
| P6 | Vic | Government | Urban planner | 15−20 |
| P7 | Vic − NSW | Private Industry | Consultant Planner | 5−10 |
| P8 | NSW | Private Industry | Urban Planner | 10−15 |
| P9 | NSW | Government | Planning Engagement | >20 |
| P10 | Vic | Utility Provider | Land Development | >20 |
| P11 | Vic | Government | Building and Planning | 15−20 |
| P12 | NSW | Government | Planning Advisor | 5−10 |
| P13 | NSW | Government | City Strategy and Analytics | >20 |

Table 2 Overview of participants in the multiple case study

The recorded interviews were transcribed for a qualitative data analysis adopting a three-level approach (Muller & Kogan, 2010):

1. Open coding: This level considered simple descriptive labels for concepts in the interview transcripts and initial grouping of themes under TOE framework;

2. Axial coding: in this level the relationship between the themes and next-level analysis was found;

3. Selective coding: This level is more focused on critical axial codes and cross-case analysis.

A case study protocol (CSP) was developed to ensure the rigor of this study and the validity and reliability of the emerged constructs. The CSP steps followed three phases of the Eisenhardt framework explained in (Maimbo & Pervan, 2005):

**Phase 1: Model development**

A problem statement and conceptual model were developed through a comprehensive literature review. The literature review employed an exploratory search approach using Scopus, and web of Science search engine. This approach was adopted to cover many relevant publications (Kurnia et al., 2015). Several search terms used include "ePlanning", "eGovernment", "Planning Support Systems", "Planning Support Science", "Digital Transformation" and "Smart Planning". More than fifty papers about topics related to digital transformation and urban planning were identified. The literature data was saved in the Mendeley reference manager. After further screening, twenty five highly relevant papers were selected to be included in this study.

In addition, a research model was developed, and a research instrument was created (See Appendix **A**). Next, initial contact with potential participants was established. Finally, a snowball sampling method (Frey, 2018) was adopted to reach an appropriate number of the sample (6 in each case study) (Maimbo & Pervan, 2005), and execute the data collection. This sampling method uses referrals made by experts who share particular interests and expertise related to this research topic. As a result, thirteen out of seventeen interview requests indicated their willingness to participate in this study.

**Phase 2: Model testing**

This phase involves data collection and data analysis . For the data collection, a triangulation of sources method was adopted to examine the consistency of different data sources (Patton, 1999). For this reason, the literature was used from different timeframes (2006, 2017, 2021) to compare if there is a divergence in the influencing factors and challenges mentioned. Then, the literature review data were compared with the interview data to examine the divergence rate. Finally, different views on the same topic were compared by selecting various experts among respondents.

**Phase 3: Model refinement**

The comparison of the resulting model with the existing models and the literature helped refine the model to be robust and generalisable.

**4. Overview of the Case Studies**

Two case studies involving Victoria (Vic) and New South Wales (NSW) are selected based on their planning systems and rules. In Victoria, the rules determine how to use a land and what to build on it are defined by the Planning and Environment Act 1987 (Planning and Environment Act), the Building Act 1993 (Building Act), and the Subdivision Act 1988 [49]. One of the main challenges in the Victorian planning system is the difference in planning schemes (rules) across their municipalities. Recently, the Victorian Government has reviewed the planning and building process and provided recommendations to continue the Smart Planning program that started in 2016. One of the outcomes of the Smart Planning program was developing a new digital planning system and associated tools to facilitate access to 70,000 pages of planning rules online. However, the government recommended that new digital initiatives should bring other planning processes, such as development assessment, online and analyse the planning system's performance (Victorian Government, 2021).

The NSW's planning system is based on the Environmental Planning and Assessment Act 1979, amended, and simplified over the last 40 years (NSW Government, 2021). The NSW planning portal is their current initiative, which hosts more digital planning services than Victoria. For example, the NSW's ePlanning Program[2] is currently working across all state's councils. In addition, the NSW government uses APIs (Application Programming Interfaces) and streamlined data exchange between ePlanning Digital Services and IT systems of the stakeholders, including the councils and private industries. Therefore, the planning rules and digital maturity levels in both cases are different.

---

[2] https://www.planningportal.nsw.gov.au/spatialviewer/#/find-a-property/address

## 5. Study Findings

A variety of perspectives were expressed in identifying the current and future influencing factors and challenges for the digital planning process. By categorising the most important themes, seven distinct groups could be identified, which broadly reflect respondents' concerns. Those themes are "technical", "data", "value and cost", "planning business", "regulation support", "resources", and "culture" (Figure 2).

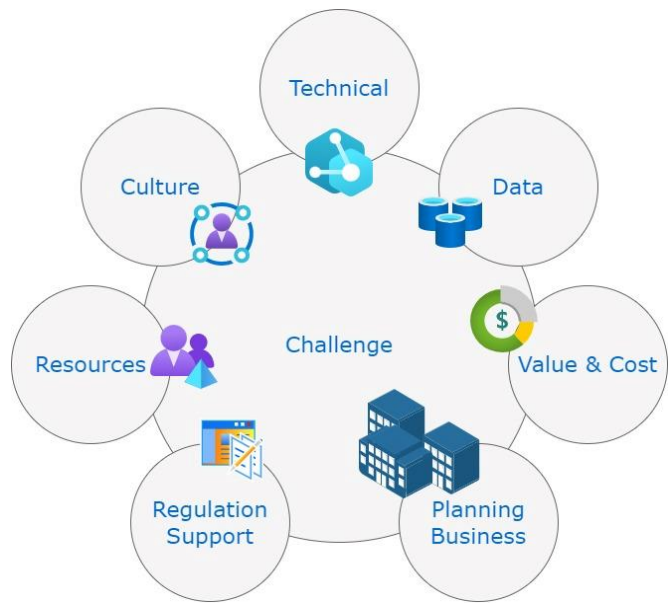

Figure 2: Current and Future Influencing factors and challenges for the digital transformation of urban planning

*Technical*

A common view amongst interviewees was that many technical issues need to be considered for the digital transformation. At the country level, the maturity level of technological capabilities among councils is *"piecemeal"*(P4). While some major cities made the high-tech applications available to their clients, the regional and rural councils are yet to implement advanced technologies (P4, P7).

One of the major concerns was the *"lack of interoperability"* with other sections inside the organisation. For instance, P12 and P13 indicated that strategic planning and urban design are more connected to the external system than their organisation. Another interviewee, who is a spatial technology expert, reports that their internal system has not been updated for a long time, and *"ad hoc"* interventions created a *"giant mess"*(P5). Although according to P3 and P4, *"there isn't one size fit all decisions"* and a *"com- bination of 2-3 out of box tools"*(P3) required to address the planning tasks. This highlights the *"lack of a common platform"*(P6, P10, P11) for all stakeholders. Moreover, one of the participants referred to the *"lack of standards"* (P7) for data exchange when working with associated authorities. Confirming this issue, one respondent indicated that while the Victorian government's pilot projects are helpful to test the idea of digital modernisation in urban planning, there is uncertainty *"for the further rollout"* (P2) of the system, including connecting with other organisations.

*Data*

In general, the participants believed that, while centralising the planning data at the state level is welcoming, the lack of reliable and accurate data is a big challenge in the journey for digital transformation in planning sections. For example, one interviewee said:

*"I think data accuracy is one of the big considerations. So for example, there are rules in the planning scheme that relate to the size and area of your land. So we need to have the size and area of the land accurate. So that we can say. . . , site coverages can't be more than 50% of that or it's less than 300 square metres. And so many developments require a permit, those sorts of rules are predicated on accurate data … I think that probably gets worse when you move from the administrative layer, like the cadastral layer, to the built form layer"* (P1)

Another respondent included:

*"I think there is actually a distinct lack of appreciation in skill across council for providing quality data"* (P5)

The participants explicitly referred to the lack of *"data standards like BIM"* in most government and private organisations. One participant thought that each organisation might prefer to have their data model based on their core business and activities. However, using *"standards"* will guarantee the interoperability of private firms with authorities (P7). Moreover, P4 and P9 referred to the data format and explained that many councils still use *"PDF as map"*, which should be converted to other digital formats to integrate with other data. There were some concerns about current data available in local governments. P5 highlighted that the data in its current form could only *"answer the basic questions to (e.g., overlay)"* the planners, which is a limitation. Adding to this point, P2 and P7 raised the issue of the lack of real-time data in city and neighbourhood levels, such as environment, and traffic, which can help better analysis and decision making. Therefore, P7 emphasised that *"how can we rely on modelling outputs?"*

*Value and Cost*

In most cases, the participants agreed that *"some local governments (Councils) don't see value"* (P4, P6, P7, P12, P13) in investing in planning technologies. This is because many local governments are small and *"finance"* is a barrier to adopting sophisticated technologies (P4, P11). Other respondents explained that *"planning is one of many tasks"* in local governments, which is not a priority (P2, P6).

However, when it came to implementing digital infrastructure, the financial value was underestimated. For example, one interviewee reported that after moving to the planning portal in the NSW case, the government faced an unexpected volume of data and started charging the users to cover the cost of data warehousing and storage (P9).

*Planning business*

The respondents alluded to the notion that while most planners are supportive of digital transformation, there is a general fear that technologies may not fully support their day-to-day activity (P4, P6, P8). One interviewee, who is an urban planner, said: *"we're more of . . . relying on our own expertise"* (P7), rather than digital tools. This view was echoed by other respondents, who also emphasised on the *"complicated"* planning scheme and tasks (P1, P3) and *"multi-component"* aspect of planning data (word and map) (P1). The respondents believed that current technological capabilities could not connect and adequately support planners for data integration and decision making. Therefore, due to these complications and multi-component aspects, currently, the *"planning process is predominantly manual"* (P5, P6, P7, P8). Two respondents related the manual and complicated planning tasks to the lack of *"planning systems harmonisation"* across councils in Victoria (P1, P9). Another challenge is the requirements of different audiences, who need a different set of tools and technologies to work with (P3). For example, the transport group requires modelling of total mobility, whereas the economic team is looking for supply-demand analysis.

Another concern was the *"disconnection"* between *"planning strategy and IT strategy"* in the organisations (P6, P12). The respondents also mentioned the lack of support from IT due to a limited understanding of planning business requirements. In one case, the participant explained that: *"I don't have an IT bone in my body but just had to learn and get on with it because the support wasn't there"* (P6).

In opposite, another interviewee believed that moving towards digital is not *"planners' core business"* (P7). Instead, they choose the easiest, open-source, and cheapest tool to communicate their plans with the clients. A public planning service (P6) and a private planning firm (P7) express these two different views.

One respondent highlighted the challenge of business-to-business interaction and information exchange. P11, who was recently involved in a digital transformation journey in the planning process of a council, explained that other organisations were unwilling to work with their new system. Therefore, most of their inter-organisational data exchange remained traditional. One of the significant considerations leading to this challenge was the definition of privacy in different organisations. For example, while the planning scheme allowed them to disclose some information to the public, the department of transport was unwilling to let a certain type of information be publicly available.

*Resources*

This theme refers to planners' and other stakeholders' digital literacy and skillsets. The interviewees consistently indicated that the PlanTech and digital skills are in infancy in the Australian context. In general, the respondents stated while there is a *"digital leadership"* (P1, P4) among the executives and high-level management, the lack of digital skills among planners, and specifically at the council level, is a big concern (P1, P2, P3, P4). In addition, the lack of digital literacy limited the interaction of planners, IT, and GIS teams (P3, P12, P13). P7 highlighted the same challenge and suggested:

*"the thing we really want to see appear is that planners understand the tech well enough that they can work alongside PlanTech software engineers and developers to ensure that the tools are like fit for purpose for planning"* (P7). Moreover, a discussion about the role of the market as a driving force for digital transformation provided new insights. For example, one interviewee believed that vendors have not been supportive of their transformational journey by saying: *"They're not the most supportive or proactive vendors in the world"* (P6). Another interviewee echoed this issue in the procurement process for upgrading their GIS system and included: *"there were a couple of particular GIS vendors who didn't choose to submit a response, which surprised me"* (P5).

*Regulation Support*

The respondents, on the whole, explicitly indicated that despite the political support,

a "strong regulatory support" towards the digital transformation of planning is lacking (P3, P4, P6, P7, P8). However, a small number of respondents had an opposite view as in the NSW; the state government mandated using Planning portal for all councils in March 2021 (P10).

*Culture*

The theme of culture emerged from discussion around willingness to commit and learn new processes and technologies. P9 refers to the planning portal mandated by the NSW government and highlights that while it has made the planning process more manageable, it is not connected to the other sections in their organisation:

*"We find it's been quite difficult to get in time investment from other people that are used to a way [platform] and they see our portal as a distraction and just another piece of Software that they've got to get used to"* (P9)

The interviewees provided examples demonstrating that *"willingness to learn"* is not similar in the entire organisation. For example, P6 explained that the supporting staff, including arborists and transport engineers, were *"less interested"* in changing. P9 and P12 highlighted the same challenge and mentioned that "not everyone was excited" in their organisations' digital transformation journey. Furthermore, P1 and P8 believed that even though everyone embraces innovation, the planners express their fear of *"eroding professional discretion"*, which is the case in discretionary planning systems like Victoria.

## 6. Discussion

The present study was designed to determine the influencing factors and challenges of digital transformation of urban planning in Australia. Through a literature review, twenty-six factors are categorised in technological, organisational, and external environmental dimensions (Table 3). This study revealed that the most significant influencing factors are the organisational and external environmental dimensions rather than technological ones, confirming the findings of several previous studies (Geertman, 2006; Goodspeed & Hackel, 2017).

Although there is some concern about the lack of regulatory support for the adoption of digital technologies in planning systems (Daniel & Pettit, 2021); this study found that in the case of NSW, the governments' commitment and actions towards standardisation of planning systems and rolling out the ePlanning platform is appealing. This observation is particularly relevant to the recent NSW's step-wise transformation of the traditional development approvals process to the digital environment, which presents the first step towards more advanced digital transformation. In contrast, the Victorian government is yet to provide a fully functional ePlanning system to the community. This is due to the fact that while there is a relatively high political backing in the digital transformation process at the national and state levels, the planning regulation reform in the case of Victoria is slow.

| Theme | Challenge | Literature Evidence (N=25) | Empirical Evidence (N=13) | Adjusted (A) / New (N) |
|---|---|---|---|---|
| **Technical** | Citizen-centric platform/ user friendliness of tools | 10 | 4 | - |
| | Lack of spatial and planning technology standards | 4 | 6 | - |
| | Lack of open-source tools | 6 | 1 | - |
| | Weakness of technical support and preparation | 4 | 2 | - |
| | Systems are developed in silo | 4 | 5 | - |
| | Maturity level | - | 3 | N |
| | Lack of systems interoperability | - | 4 | N |
| **Data** | Lack of planning data standards | 6 | 4 | - |
| | Data availability and usability / Open government data (OGD) | 7 | 4 | - |
| **Value and Cost** | Scarcity of financial capital | 5 | 5 | - |
| | Overlooking the priority of digital investments | 4 | 7 | A |
| | Perceived cost of technology for planning | 3 | 5 | A |
| | Value proposition for planning technologies | - | 3 | N |
| **Planning Business** | Lack of IT-Planning strategy alignment | 6 | 7 | A |
| | Planning complexity | 7 | 5 | - |
| | Busines-to-business interactions | - | 1 | N |
| **Resources** | Digital literacy among planners | 6 | 5 | - |
| | Capability for adapting to change among planners and other associated disciplines | - | 3 | N |
| | Lack of relative skillset among planners | 6 | 5 | - |
| | Market and vendors | - | 3 | N |
| **Regulation Support** | Political support | 8 | 5 | A |
| | Regulatory support | 5 | 5 | |
| **Culture** | Lack of inter-organisational collaboration | 8 | 7 | A |
| | Culture of communication inside and between organisations | 3 | 4 | A |
| | Embracing the innovation among all stakeholders | 7 | 3 | A |

Table 3 Summary of results and comparison with the literature

The culture of inter-organisational collaboration also affects the success and effectiveness of digital technologies in planning systems. This study in- dicates that both NSW and Victoria governments started centralising the databases and digital platforms to harmonise the data integration, and facilitate a better collaboration across the local governments and service providers (e.g. department of transport, utility providers, department of environment). However, the study observes significant differences across local councils, in the two study areas, regarding their digital capability and IT support, leading to the prevalence of technological complexity levels to facilitate a smooth transformation towards digital technologies amongst all stakeholders. This is true especially among the rural governments, where there is a scarcity of budget and a lack of priority for digital planning reform, confirming the findings of Musakwa Moyo (2020) in South Africa and Criado-Perez et al. (2021) in Brazil. This matter further impacts the mismatch between IT-Planning strategy alignment, creating an intra-organisational collaboration challenge.

This study further indicates that the lack of capability to adapt to the change affects inter and intra organisational collaborations. This finding conforms to Agius et al. (2018) and Christensen and Lægreid (2020). The results suggest that adapting to organizations' digital technology is related to the cultural and technological aspects. From a cultural perspective, the study highlights that there is more resistance to innovation among supporting staff, including arborists and traffic engineers compared to the planning professionals. From a technological perspective, it is found that digital technologies are developed in a silo, and the exchange of data between systems is problematic both inside and between organisations. This observation is more dominant in the Victorian case as compared with the NSW. The difference between two cases may be explained by the fact that ePlanning system is functional in NSW and the state government maintains and improves the system. Nevertheless, other relevant disciplines such as digital platforms for land sub-division and land registry are well integrated with state, local government, and utility providers, in both Victoria and NSW cases.

The study underlines the significance of the two theoretical dimensions of the PSScience: governance and instrumentation. In Australia's case, contrary to some European jurisdictions (Tomor & Geertman, 2020), the smart cities program was initiated by the federal government and stimulated the digital transformation of urban planning. However, there is a lack of a specific federal strategy or a roadmap towards digital transformation. This limitation impacted the lack of collaboration between different organisations.

Using the governance lens of PSScience, this study found that while the digital leadership, commitment, and literacy among executives and high-level management is well-established, the culture of communication and collaboration among different stakeholders remains a critical challenge. The councils have already started their digital transformation journey. However, surprisingly, despite their expectation, they have not received enough interest and support from different vendors. This leaves the organisations with limited options for upgrading, implementing, and maintaining digital infrastructures. A potential explanation for this is that Australia is a relatively small market for the high-tech industry, resulting in the lack of support and proactive presentation of vendors.

From an instrumentation perspective, the study observed that the lack of IT – Planning business strategy alignment led to the challenge of lack of systems interoperability, which results in the overlooking of the digital planning value. As an example, the planners in both case areas cannot use new types of data such as real-time sensor readings about the environment and traffic generated by other departments in the same organisation or other organisations in the government. This lack of access to real-time data limits the contribution of urban planners with little data-driven and evidence-based decision-making.

Interestingly, the digital maturity model is absent within the Victorian and NSW urban planning sector. The planning industry and the government do not seem to actively advocate and support digital technologies' systematic adoption and penetration. While the peak body started developing principles and best practices for adopting PlanTech, there should be an academic-industry partnership for awareness, upskilling, and capacity building. Such a partnership provides a platform for training digital leaders at different levels of professional, managerial, and executives in the urban planning sector. This study observed a growing interest in digital reform in both state governments. However, there is little collaboration between industry and academia for defining future skills requirements and planning for strategic change in digital education, training, and awareness programs.

These findings raise intriguing questions regarding the essential role of regulatory reform and developing standards specific for digital technologies and data models in planning activities. From a regulatory reform perspective, pilot projects are necessary to inform the policymakers about the extent to which the planning scheme and other regulations are ready to accept the new business models incorporating digital technologies. The experience of pilot projects in Victoria contributed to understanding regulatory reform requirements. However, more specific projects, such as automating the planning tasks through adopting emerging technologies (e.g., AI, ML) in collaboration with academia, industry, and regulatory authorities, will provide further opportunities for data-driven and evidence-based reforms. The study also highlighted an important implication for developing technological and data standards and interoperability. The observations suggest that the technology and data harmonisation standards should be vendor and regulation agnostic and accommodate the new type of data and analytics for more inter-organisational collaborations.

In sum, this study indicates that while the maturity of digital planning adoption is different across the two study areas, the Australian planning industry generally embraces digital transformation planning services across the government. By considering the organisational factors, the study enriched our understanding of the essential role of capacity building and improving digital literacy among planners to better empower digital transformation advocates to communicate the value proposition of planning technologies. Accordingly, better communication of digital planning values will lead to mutual understanding and better alignment of IT-Planning strategies. Therefore, the study demonstrated how the influencing factors and underpinning challenges for digital transformation in Australian urban planning can be better perceived by adopting the TOE framework.

## 7. Conclusion

This study was undertaken to explore the challenging factors under the influence of technological, organisational, and external environmental dimensions for the digital transformation of urban planning in Australia. The study employed a multiple case study approach involving government and private industries in Victoria and NSW. It was found that generally, the organisational and external environmental factors are the most significant challenges for the digital transformation journey in the urban planning systems of both Victoria and NSW governments.

The findings of this study enrich the current literature by offering a better understanding of the challenges of digital transformation in a highly manual and paper-based (PDF) planning system. Therefore, this study complements previous findings and provides new directions for future research. Among the organisational factors, the value proposition for planning technologies was found to be a significant challenge. As a result, most of the local governments in Victoria and NSW consider planning services as less priority. As compared to the Victorian local governments, due to the availability of ePlanning in NSW's local governments the planning technology values can be better perceived.

In addition, the study discovered that one of the critical organisational challenging factors is the lack of the IT-Planning strategies alignment in both study areas. One potential explanation for this is that the planning specifications and requirements are not often clearly communicated in the process of digital strategy development. Therefore, there is a need for capacity building and improving digital literacy and digital leadership among planners, which will empower them to better communicate the digital planning values and potential outcomes.

Another significant contribution of this work is the emphasis and insights provided on digital transformation challenges in the context of multiple planning systems. As planning schemes vary across NSW and Victoria, factors influencing the technological transformation strategies are provided.

The findings support the organisations in adopting regulation agnostic, interoperable, and standard digital technologies and data models. In addition, the general planning industry can leverage these findings, which suggest the potential role of integrating with other sectors such as AEC in addressing planning complexity challenges. This aspect also improves understanding of the overall digital transformation in public sectors related to urban development. Using the PSScience factors under the TOE framework helped better understand the interplay of the above-mentioned challenging factors, representing a genuine contribution to this study. The findings will be of interest to the Australian professional urban planning industry as it offers new insights into the current application of digital technologies. It is critical to consider that while the Australian market is small for high-tech vendors, an investment in open source and open platforms will support the planning community. This understanding is especially beneficial for the professional bodies and industry to partner with academia to educate and train the current and future workforce and augment the digital skillsets and literacy among planners. Identifying challenging factors also help organisations evaluate their positions for moving towards digital transformation and develop an appropriate roadmap.

The generalisability of these results is subject to certain limitations. For instance, the study only considered the two states of Victoria and NSW in Australia. Other states (e.g. South Australia) and councils (e.g. City of Hobart) have already extensively developed the digital transformation. Therefore, they may have different experiences and other challenging factors. Furthermore, since the study was limited to data from only one representative infrastructure department as an external organisation, it was not possible to discuss the inter-organisational challenges in more detail. In addition, understanding the views of high-tech vendors (as an external organisation) is vital to discover the main challenges for lack of interest in participating in the digital transformation of urban planning. By understanding the role of associated organisations, end-users, and vendors in the future, we can draw valuable insights into the broader ecosystem of the urban planning industry and incorporate their challenging factors for better strategic development. Ultimately, as this study is based on multiple case studies involving two states of Victoria and NSW within Australia, the findings should be interpreted cautiously within the relevant context.

## 8. Acknowledgement

## 9. Appendix A,

| Introduction | Description |
|---|---|
| **Name, Occupation, Organisation** | |
| In what capacities and for what duration have you and your organisation been involved / concerned with Digital Transformation of urban planning system? | The planning tasks can be statutory or strategic. It is important to know in what capacity the digital transformation is mature |
| **Technological Factors** | |
| What type of technologies do you use for the urban planning task? | Availability and characteristics of technology in terms of its complexity for use |
| What type of technologies do you use for the urban planning task? | Availability and characteristics of technology in terms of its complexity for use |
| How complex are the available technologies for delivery of tasks? | Complexity of technologies are sometimes a barrier |
| How compatible are the technologies you mentioned with existing culture and values in your team? | compatibility with existing culture and values |
| Do you think it is feasible to reinvent your core mission operations, services or outcomes using the digital technolo- gies? What might be the potential risks? | feasibility of adoption in organisation with minimum risk |
| **Organisational Factors** | |
| Is the current political and executive leadership of your organisation spon- soring — that is, actively supporting — digital transformation? | Top management support can be through developing digital strategy or business strategy with focus on digital technologies |
| Is there any perception or indication for costs of implementing digital tech- nologies in your organisation/team? Is there any mechanism for securing fund? | Perceived costs as one of challenges |
| Does your team (Development assess- ment, not just the IT department) face problems with attraction, moti- vation, satisfaction and retention of skilled staff due to digital workplace challenges? | Understanding the challenge of avail- able skillset for digital technologies and future innovations |
| How would you describe the overall attitude toward change in your organisation (among employees, middle managers, leadership)? | Organisational culture |
| **External Environmental Factors** | |
| Approximately what share of your organisation's user-facing digital services supports at least one of the following identification mechanisms: Proprietary digital identification that is interoperable with services in other institutions Digital identification of another institution (such as referral authority, state or commercial entity like a bank or mobile operator) Shared digital identification or single sign-on across different public sector services | It is assumed that organisations with close relationships adopt a similar structure, even though the structure's efficiency is unknown |
| To what extent current infrastructure in your organisation supports the new digital technologies for your tasks? | Availability of infrastructure |
| How much is your organization's sector of activity (mission area, policy domain, market segment) affected by current societal pressures, political agendas and policy choices of your government? | Market demand |
| Are there any political or regulatory supports for this transformation? If so, how much is your team's digital agenda affected by recent changes to the legal and regulatory framework (local, national and international)? | Political and Regulatory support |
| **Concluding Question** | |
| Are there any other issues you would like to comment on further or that you feel we have not addressed or that we may have overlooked in this interview? | |
| Opportunities | |
| Risks | |
| Challenges | |

Table 4: Questionnaire